# Bridging the neuroscience and physics of time

*Dean Buonomano[1] and Carlo Rovelli[2,3,4]*

To appear in: *Time and Science* (ed. Lestienne & Harris)

1. Departments of Neurobiology and Psychology, University of California, Los Angeles. Los Angeles, 90095.
2. Aix Marseille University, Université de Toulon, CNRS, CPT, 13288 Marseille, France
3. Perimeter Institute, 31 Caroline Street North, Waterloo, Ontario N2L 2Y5, Canada
4. The Rotman Institute of Philosophy, 1151 Richmond St., London, Ontario N6A 5B7, Canada

## 1. Different views on the nature of time

There are differences between the way a neuroscientist and a theoretical physicist—such as the two authors of this chapter—comprehend time. For a neuroscientist, time is oriented, always pointing towards the future. We can remember the past (but not the future) and we can influence the future (but not the past). The past no longer exists, the future is open; the present is the only real moment—thus the possibility of time travel to the past is limited to fiction because one cannot travel to a moment that does not exist. For a theoretical physicist, time is more complicated: relativity does not permit an objective notion of a global present, the distinction between past and future requires thermodynamics, hence is statistical only. It is far from obvious why we remember the past but not the future, and why we can influence the future but not the past. There is a sense in which it is easier to think about the whole of spacetime as a single four-dimensional entity (the so-called block universe), in which temporal notions are a matter of perspective. Traveling back in time becomes a subject of theoretical investigation.

The striking discrepancies between the intuitive conceptualization of time in neuroscience and the fundamental physics does not pertain solely to neuroscience. But neuroscience, and psychology, find themselves in a peculiar position because these fields deal directly with the temporal nature of our subjective experiences, and thus speak directly to these discrepancies. Because of this, it is often the case that neuroscientists and physicists place the responsibility for explaining the discrepancies in these perspectives on each other's shoulders (Callender, 2017):
-Physicist: It is up to the neuroscientist to account for the subjective experience of the flow of time, if this is not accounted for by physics, as it may be a side-effect of how the brain uses information about the past to predict the future.
-Neuroscientist: Our sense of the flow of time reflects reality, and permeates all our theories of brain function and evolution, it is thus up to the physicists to reconcile this fact with the physics.

As a neuroscientist and a theoretical physicist, both working on time, we have decided to open a direct dialogue to examine if the apparent discrepancies regarding the nature of time can be composed.

We put forth that an important first step in addressing this challenge, is to disentangle several distinct problems from one another. One source of confusion is to take time as a single concept, failing to recognize its layered structure. Clarity requires distinguishing aspects of time that are rooted in different phenomena. In particular, we need to separate the aspects of temporality that are accounted for by physics, from those that can be accounted for by neuroscience. Here we attempt to partition a vast land of problems related to time: assign portions of sovereignty to the two fields, and determine which issues will benefit from a bridge between neuroscience and physics.

We have found a large agreement between us, but not complete. Hence, we separate this presentation into a first part, where we agree, and a second part where we express our remaining differences. We hope that this text could stimulate the discussion and help it moving toward increased clarity.

## 2. The sources of the discrepancies: time reversal invariance of the fundamental physical laws and relativity

We start by recalling why the theoretical physicist is led to reject the idea that the commonsense view of time could remain valid outside a limited domain. There are three reasons for this. We only list them here, without discussing them. We refer to the abundant literature on each of them (e.g., Zeh, 1989; Price 1996; Carroll, 2010; Callender, 2017; Rovelli, 2018, 2019).

(i) All elementary mechanical laws of nature that we know are invariant under reversal of the direction of time—a principle referred to as charge conjugation, parity inversion, time reversal (CPT). These include classical mechanics, electrodynamics, quantum theory, general relativity, quantum field theory, and the standard model. For example, Newton's equations are time invariant because if they permit a certain phenomenon, they also permit the time reversed version of the same phenomenon (i.e., running backwards in time). It follows that the manifest time orientation of the world around us can only be a macroscopic phenomenon that is accounted for in terms of the distinction between micro and macrophysics, and caused by low entropy in the early universe.

(ii) Relativity is incompatible with an objective notion of a global present—for example, it does not make sense to ask what an astronaut on a distant spaceship is doing "now".

(iii) Many current attempts to write a quantum theory of gravity, such as loop quantum gravity, do not utilize any time variable at all in their basic equations.

We shall not be concerned with (iii) here as it remains an open question. (i) and (ii), on the other hand, are established science. The fact (i) implies that the arrow of time is not part of the basic grammar of nature, rather it is a complex probabilistic macroscopic phenomenon. The arrow of time is a consequence of the increase in entropy imposed by the second law of thermodynamics and the low initial entropy of the universe. The fact (ii) implies that the distinction between past, present and future must be more subtle than in the commonsense view of temporality. With respect to here and now, many events in the universe are neither past nor future.

Much as in physics, time plays a special role in neuroscience. The brain is an inherently temporal organ because in many ways its primary function is to learn from the past in order to best predict the future. The degree to which animals predict the future is reflected in their ability to find food, anticipate danger, reproduce, and ensure the survival of their young. As a result, the brain of all animals endows them with the ability to form and retrieve memories, tell time across a wide range of scales (from microsecond differences in interaural time delays used for sound localization) to the hours tracked by our circadian clock (allowing animals to anticipate environmental changes imposed by the rotation of the Earth) (Buonomano, 2007; Buonomano, 2017). Furthermore, some animals, most notably humans, have the ability to consciously revisit the past and project themselves into potential futures—an ability referred to as mental time travel (Suddendorf and

Corballis, 1997). Additionally, humans have the ability to conceptualize time and communicate ideas such as "I'm looking forward to seeing you next month". Finally, human experience is anchored by the distinct feeling of the flow of time, that each present moment is continuously slipping into the past while opening the gates to the future—a future that we strive to control.

Time thus permeates many of the questions neuroscientists and psychologists study. For the most part, these questions are answered with little regard to the physics of time. Instead, neuroscience relies on a commonsense understanding of temporality in which time is flowing, oriented, the past is fixed, and the future is open.

The question is, are neuroscientists and physicists talking about the same topic when they talk about time? As we shall see, the answer is, to some extent, "no". But this may not necessarily mean that there is an inconsistency.

### 3. Two extreme positions: global presentism and static eternalism

Two extreme formulations of the views on the nature of time that are presented above sometimes go under the names of presentism and eternalism. Both terms refer to generic designations that are interpreted differently by different authors. Here we give perhaps a bit caricatural account of both, and thus qualify both by referring to them as *global presentism* and *static eternalism*. We do so to stress what is *wrong* about each in their extreme versions (Rovelli 2019).

According to global presentism, only the present is real, and under the most extreme view there is only a single real, and absolute, objective present, extended all over the universe. Time is flowing, and change is real. The past is fundamentally different from the future. Under global presentism, if relativity theory fails to describe this objective global present, it must be because it fails to capture a fundamental aspect of reality. In the words of Roger Penrose, for instance: "It seems to me that there are severe discrepancies between what we consciously feel, concerning the flow of time, and what our ... theories assert about the reality of the physical world". (Penrose, 1989)

According to static eternalism, the past, present and future are equally real, and the universe consists of a static four-dimensional "block". Here there is no objective present, no flow, no change, nor actual dynamics. The passage of time is an illusion, and in a sense the future is "already there". In the words of Herman Weyl, for instance: "The objective world simply is, it does not happen. Only to the gaze of my consciousness, crawling upward along the world line of my body, does a section of the world come to life as a fleeting image in space which continuously changes in time" (Weyl, 2009/1949).

Both these extreme views of time are inappropriate to describe nature.

Presentism in the form pictured above is manifestly incompatible with relativity theory. The argument that relativistic theories miss a fundamental aspect of reality, which would instead be directly perceived by our ourselves, cannot be correct, because it implies that there is something in our behavior not accounted by relativistic theories, contrary to all current evidence. All irreversible phenomena that we observe *are already* very well accounted for by their macroscopic description based on time-symmetric fundamental laws, and progressively increasing entropy from a low entropy state. Postulating something else is unnecessary, and it would be far from clear how that something else would reconcile the discrepancies with global presentism.

On the other hand, to say that there is no flow of time, no change, nor actual dynamics, and that the 4-dimensional universe is static is to fundamentally misunderstand physics. Physics is not the description of static entities: it is the description of processes. The 4-dimensional universe is not an entity, it is a process. Physics is about events, about change. What spacetime describes is a complex network of changes, not a static 4-dimensional block. Upon careful examination, to say that 4-dimensional spacetime is like a "block" is to imagine an additional external time variable, in which the 4-dimensional universe is remaining static. But there is no additional external time variable. The 4-dimensional universe is our map of a multifaceted set of changes. Specifically: to say that the future is "equally real as the present" is an unnecessary redefinition of the word "real", that conflicts with our commonsense use of this word. Relativity does force us to use this word with more caution, but it does not force us to this a-temporal use: we can still say that something is "real here and now", without for this having to say that the "the future is real now".

**4. The key towards a solution: time is a multilayered concept**

The trouble with the extreme views described above is that they both take "time" as a single monolithic notion, failing to incorporate the fact that "time" is a multilayered concept. First, while some properties of the commonsense view of time—such as a global present—do not extend to arbitrary large scales, this does not imply that all aspects of temporality suddenly disappear as soon as we go to relativistic physics. Second, the absence of some properties of time in fundamental physics does not mean that these properties are illusory: it only means that they are not universal: they are real, but they appear only within some approximation.

Some aspects of time that are familiar to us are rooted in specific contingent aspects of the physical domain to which we have normal access (macroscopic, small relative velocities and small spacetime curvature), others are not. To get some clarity on the nature of time, we have to break it apart in its various layers. And as a first step towards unfolding the different layers of time we distinguish between the following features of time:

(i) *Events, processes, and change.*
(ii) *Distinction between timelike and spacelike intervals.*
(iii) *Global and local simultaneity surfaces.*
(iv) *Distinction between past and future.*
(v) *Traces of the past and the possibility of affecting the future.*
(vi) *The special role of the present.*
(vii) *Memory and prediction.*
(viii) *The conceptualization of time and mental time travel.*
(ix) *The subjective sense of the flow of time.*

Let us discuss them one by one.

(i) *Events, processes, change.*
All of physics is a description of processes, events, and change. This is not questioned by general relativity (nor by quantum gravity). Therefore, there is no reason to replace the picture of a dynamical world with a static world.

(ii) *Distinction between timelike and spacelike intervals.*
In general relativity, given two distinct events *A* and *B*, whether their separation is timelike (when events A and B can directly share information) or spacelike (when they cannot directly share information) is a physical objective fact. Therefore, the distinction between spatial versus temporal intervals is encoded in fundamental physics (at least as long as we disregard quantum gravity, as we do here), and therefore there is no ground for any claim that "time becomes entirely like space" in general relativity.

(iii) *Global and local simultaneity surfaces.*
Objectively defined global simultaneity surfaces are not defined in general relativity. Therefore, the notion of a global or absolute present is incompatible with the current understanding of the universe. No empirical observation is contradiction with general relativity; thus the notion of global present is not needed to make sense of the world we observe. The fact that our intuition is attached to a notion of global present is bias of our intuition, which evolved to be adapted to our local environment. The universe is dynamical, but without a global present. However, because we live in a region of the universe with low curvature and the velocity of most objects we deal with is negligible with respect to the speed of light, the physical phenomenology at our scale is well described by Newtonian mechanics—and this gives a well-defined local approximation to a global present.

(iv) *Distinction between past and future.*
Due to the fact that we interact with a number of dynamical variables that is small compared to the number of degrees of freedom dynamically related, and to the fact that the entropy defined by these variables was low in the past universe, the physical phenomenology at our scale is well described by thermodynamics and definitely time-oriented. In this sense the distinction between past and future is objective.

(v) *Traces of the past, and possibility of affecting the future.*
Due to past low entropy, system separations, and long thermalization times (metastable states) that characterize our environment, the entropy gradient produces abundant traces of the past (Rovelli, 2020a) and allows the existence of agents that influence the future (Rovelli, 2020b).

(vi) *The special role of the present.*
At any specific spacetime location, a notion of local present is defined indexically. Since there is nothing outside the universe itself, every map of the whole of physical spacetime is necessarily itself located in spacetime, and at each time of its world history there exists a "special event", which is the fixed point of the map that sends spacetime to its representation (Ismael, 2007). Hence the local notion of "present" is always defined: it is the "now" at that event. It splits the rest of the universe into a future, a past and a spacelike separated region, which are defined objectively, but *relatively* to that "here and now".

(vii) *Memory and prediction.*
The existence of traces of the past produced in the macroscopic world by the entropy gradient permits in particular the biological utilization of information and the utilization of the information about the past by the neural system of animals. Specifically, the traces of the past include our memories, which are used to predict, and affect, the future. Hence neural systems develop in a context where there is a well-defined time variable, a well-defined time orientation, and an abundance of traces of the past, including those on evolutionary and neural time scales.

(viii) *The conceptualization of time and mental time travel.*
Humans, and perhaps some other mammals, evolved the ability to use memories to consciously recreate past experiences, and perform complex simulations of the future—that is, to engage in mental travel. Indeed, our past episodic memories seem to anchor our ability to engage in future oriented mental time travel, for example, amnesic patients exhibit an impoverished ability to imagine future scenarios (Hassabis et al., 2007; Race et al., 2011). The ability to conceptualize time and engage in mental time travel is hypothesized to have been co-opted from the neural circuits that allowed animals to perform the much more basic and fundamental task of navigating through space (Nunez and Cooperrider, 2013; Buonomano, 2017).

(ix) *The sense of the flow of time.*
The brain works, in part, by accumulating traces of the past (memories), computing possible futures, and using higher-level computations to govern an organism's behavior. These computations, of course, happen in time, thus our conscious perception of external events are delayed in relation to when they happened. Furthermore, our sense of the linear flow of time is an illusion in the sense that later events can retroactively alter our conscious perception of earlier events (Kilgard and Merzenich, 1995; Goldreich and Tong, 2013; Buonomano, 2017). From the perspective of physics, entropy gradients, traces of the past, and the macroscopic under-determinacy of the future, underlie the brain's ability to produce a subjectively vivid and richly structured "flow of time". This flow is subjective both in the sense that it is part of each individual's conscious experience, and in the sense that can be decoupled with clock time—i.e., the rate of the flow of time can speed up or slow down depending on emotional state and context (van Wassenhove, 2009; Arstila and Lloyd, 2014).

**5. Partitioning the land.**

In light of the discussion above, it becomes clear that understanding the phenomenology of experiential time requires a delicate interplay between physics and neuroscience. Physics provides:

(a) the understanding of the *general* temporal structure of the world. See (i-ii-iii) above, namely the notions of process, change, dynamical correlations, and so on;

(b) the justification for the existence of *additional* properties of temporality that become significant in the appropriate approximations and hold in *specific* domains relevant for biology. See (iv-v-vi-vii) above, namely the physical basis for a preferred time variable, for the arrow of time, and for the notion of a local present.

Neuroscience *builds on the conceptual structure that is appropriate to describe these domains*, which include oriented evolution and a distinction between past present and future. Neuroscience has historically taken presentism as its starting point, since our subjective experiences are obviously consistent with it, and present, past and future are clearly different in the conceptual structure of the discipline. The important point here is that in order to anchor the conceptual structure of the neurosciences these features *do not need to be universal*: it is sufficient that they hold in the relevant domain and at the relevant scale. They do, as physics shows, therefore there is not necessarily a conflict.

Neuroscience builds on the existence of macroscopic traces and on the openness of the macroscopic future produced by the thermodynamic arrow of time. The second, in particular, underpins the possibility of our experience of being "free to choose", since different macroscopic futures are compatible with the same macroscopic past, choice depends on what happens in the organism. These notions permit the neuroscientist to study the brain as a computational device that uses the past to predict the future. The presence of traces of the macroscopic past and the openness of the macroscopic future justify the language of a past that "has happened hence is fixed", and so on.

By analogy, these concepts are a bit like the notions of "vertical", "up" and "down: they are absent in Newtonian physics but become relevant on Earth. We understand the full story for which the notions of "vertical", "up" and "down" are well defined on Earth, in spite of the fact that they are not well defined everywhere in the universe. Similarly, we largely understand how the notions of a preferred oriented time variable allowing irreversible processes can be well defined in the domain where the biosphere develops, even if they are not universally defined. On the basis of these elementary notions, neuroscience explores the rich manner in which these temporal structures are utilized by the brain in its basic functioning, and is the position account for the full richness of our experience of the passing of time, memory, anticipation, and so on.

The research land seems therefore quite clearly partitioned.

Questions that pertain to physics, that have no meaning in neuroscience include:
- Why are there traces of the past but not the future?
- Why was entropy low in the past?
- Why macroscopic phenomena strongly time oriented?
- What is the nature of time beyond the non-relativistic approximation?

Questions that pertain to neuroscience, and have little meaning in physics include:
- What is the subjective "now", and what time window does it reflect—i.e, what is the "width" of subjective now?
- What determines the sense of the speed at which we perceive time to pass?
- How does the brain use memories to allow us to engage in past and future oriented mental time travel?
- How does the brain tell and represent time across all the time scales necessary for survival and cognition?

Our intuition about temporality is colored by the complexity and richness of experiential time. The phenomenology of our experiential time is closely entangled with problem of consciousness. Indeed, it is hard to conceive of consciousness in the absence of the subjective experience of change and the flow of time. The subjective phenomenology of time include the sense of time flowing, the sense of anticipation, nostalgia, longing, a sense of a given `speed' at which time passes, the fluid and fleeting feeling of the present dissipating into the past, the ability to predict the future and the resulting curse: the anguish of impermanence. It is difficult for us to disentangle the different components of temporality, and properly separate those that have nothing to do with our brain from those that emerge from our brain. Experiential time is too much of a constitutional part of our conceptual structure to easily correct it to account for the proper conceptual structure needed to understand larger domains of nature. For instance, thinking of the present as a local property, distinct from the fantasy of global present, can be counterintuitive, because the question what time is it now on Andromeda seems like it should be a perfectly valid question to ask. We believe that these difficulties are at least in part of the same kind as the difficulty that our ancestors had, overcoming the notion of a flat earth and accepting the idea that at antipodes people could are in a sense upside down without feeling that way.

A common mistake is to promote some parts of experiential aspects of time to universal features of nature and then perhaps marvel if we do not find them in physics. Understanding experiential time is a problem for neuroscience and psychology, not physics.

Yet the boundary between the competences of the two disciplines remains hazy in some domains. Since it is not easy for us to develop a good intuition of temporality in the domains which are outside the non-relativistic and thermodynamical approximations (even more so, outside the approximation that disregards quantum gravitational effects), we tend to be confused about the extent to which our subjective experience of the flow of time and the perceived specialness of the present, are products of the brain's architecture, and to which extent it reflects the true aspects of all things expressed by the dynamical equations of physics.

For this reason, it is essential, in our opinion, to sharpen not only our understanding of fundamental physics, but also our understanding, which is rapidly progressing, but still insufficient, of the way our brain deal with time and give us the experience of the passage of time.

## 6. Differences.

So far, we have managed to express a common perspective. It is now time to separate our voices and say where we still find remaining differences in our understanding of time.

DEAN. As a neuroscientist I favor the notion of what might be called *local presentism*. Consistent with relatively there is no absolute present, and the notion of *now* is inherently local. While 4D spacetime coordinates are needed to describe dynamical systems and the laws of the universe, the temporal dimension is not a physical feature of the universe. Our intuition that the local past and future do not exist is reflects reality, and the 4D block universe that emerges from relativity, along with the "realness" of the past and future is taken as an interpretation, not a prediction, much less a testable one. Ultimately, both in neuroscience and physics, time is quantified by clocks, which are local devices—and whether the clock is of the atomic sort or the brain sort, they are confined to, and evolve according the laws of physics, in the everchanging local present.

The neuroscientist further emphasizes that, for better or worse, there is one unsurmountable bridge between neuroscience and physics: the fact that the interpretation of laws of physics are filtered through a computational device that did not evolve to understand the mysteries of the universe. The brain is a product of haphazard evolutionary design, and thus constrained by limitations, biases, and bugs. The best scientific tool at our disposal to immunize ourselves against these limitations and biases is called *mathematics*: once a theoretical physicist conjures up equations that capture fundamental truths about the universe, these equations are independent of the human mind—i.e, they can be used by computers and machines to predict the future and retrodict the past. But a problem can arise when we attempt to interpret these equations within the confines of a biological device that clearly did not evolve to make sense of, for example, the equations of general relativity. Thus it is possible that the block universe of static eternalism emerges as the interpretation that produces the least cognitive dissonance within our spatially-biased neural circuits (Smolin, 2013).

Furthermore, unlike the empirically confirmed predictions of relativity (e.g., clocks slow down at high gravitational potentials), it is important to stress that there is no empirical evidence for the block universe. Indeed, it is far from clear that there are any testable predictions that could prove or disprove the existence of the block universe (other than the emergence of a confirmed time traveler). Thus, the neuroscientist takes the position that that our subjective experiences evolved to survive in an unwelcoming world, a world governed by the laws of physics. Survival however was not dependent on grasping the physical laws on the quantum and cosmological scales—which is why our intuitions fail miserably on these scales. Many fundamental questions pertaining to time however, such as whether the future is open or if the past, present, and future are equally real, are valid across scales. Thus, our intuitions about time—which, again, evolved to help us survive in a world governed by the laws of physics—should be taken seriously. And the irrepressible feeling that the local past and future are fundamentally different from the present may reflect a fundamental truth about reality (across scales)—a truth that is not inconsistent with the known laws of physics, but rather is simply at odds with the most common interpretation of these laws after they are filtered through the human brain.

The fundamental tension between the physics and neuroscience of time is often posed in regard to whether the subjective sense of the flow of time reflects a reality about the universe. To me, however, the more fundamental intersection between physics and neuroscience pertains to the reality of the past and future, and specialness of the present. Ultimately, science is anchored in empirical evidence, and while it is an empirical fact that time as measured by a clock changes according to its speed and local gravitational potential, there is no empirical evidence to support a critical tenet of the block universe: that the past and future, physically speaking, are as real as the present. We agree that perhaps the cleanest way to see the difference between us it to ask the question whether closed timelike curves (time travel) are a theoretical possibility or not—under local presentism the very nature of time would forbid the possibility.

CARLO: For a physicist such as myself, the intuitive understanding of time cannot be extrapolated beyond the regime where it works: *macroscopic* systems *away from equilibrium*, moving at low relative *speed*, *small* with respect to the local radius of the spacetime curvature, and *large* with respect to the Planck scale. This is a vast domain that includes the biosphere. There is no necessity to assume our intuition about time to holds beyond this regime, because we have a good understanding of how the intuitive properties of time emerge *in this regime*, as described in the points (i-vii) above.

The "block universe" is a misleading metaphor because it suggests a static reality. Reality is change and the 4-dimensional spacetime of General Relativity is not static: it is a way to account for change. At each location of the universe and at each specific moment in time, we can say that the present is "real". At different times, there are different "presents" that are real, and a 4-dimensional spacetime is a listing of these distinct temporal moments. A world line, in turn is an abstract representation of a sequence of events that happen at different times. Hence there is no absolute present over and above the present naturally represented by any specific point of spacetime. In simpler words, it is right to say that now and here there exists a special present. It is wrong to say that there is a special present, without explicitly or implicitly specifying when. In the same sense, past and future do exist *now*. Hence they are defined relative to each point in spacetime.

Physics is not only measurements and math: it is also the capacity of developing new concepts. The works of Copernicus, Darwin and Einstein, just to mention major examples, have required important changes in the way we conceptualize nature. Our brain has limitations and biases, but it is remarkably flexible and capable of learning new conceptual tools. Perhaps this flexibility is a biological feature of our species. It was hard for us to accept the idea that the Earth is a sphere and people do not fall away from it at the antipodes. Hard to digest the idea that the Earth is spinning fast and yet we call the ground "not moving". Similarly, it is hard to develop an intuition for relativistic spacetime. Cognitive dissonances are unavoidable, but are welcome, because they are steps towards getting out from mental habits and learn. The empirical evidence for relativity is overwhelming, and I think that we should take its conceptual structure seriously.

Nothing in the discoveries of fundamental physics questions the appropriateness of the conceptual tools that a neuroscientist uses: on the contrary, fundamental physics shows that they are appropriate, and well founded in physics. They are appropriate to a limited domain, and grounded in approximations.

In particular, the irrepressible feeling that the local past and future are fundamentally different from the present correlates with a fact of the physical reality of the domain of which we have experience. It is consistent with the laws of physics. But it does not correlate with the physical reality outside this domain. The is no reason to believe that the intuition based on our limited experience knows more about the universe than our best physical theories.

As Dean put it, a sharp way to characterize our differences regards the potential existence of closed time-like curves and time travel. I see no reason to believe that closed timelike curves are impossible. The impossibility of "time traveling" to the past is thermodynamical, hence statistical: the future is the ensemble of phenomena permitted by entropy increase and entropy cannot increase along a closed path, because a function cannot be monotonically increasing on a circle. Our differences center on the question why do we have a sense of a time flowing from a fixed past to an open future. I think that this sense of oriented "flowing" is due to our memories and anticipations and is permitted by the physics of the specific regime in which we live, characterize by a marked entropy gradient. It is does not pertain to the most general grammar of nature.

References

Arstila V, Lloyd D, eds (2014) Subjective time: the philosophy, psychology, and neuroscience of temporality. Cambridge, MA: MIT Press.
Buonomano DV (2007) The biology of time across different scales. Nat Chem Biol 3:594-597.
Buonomano DV (2017) Your brain is a time machine: the neuroscience and physics of time. New York: W.W. Norton.
Callender C (2017) What makes time special?
Carroll S (2010) From eternity to here: The quest for the ultimate theory of time. New York: Penguin.
Goldreich D, Tong J (2013) Prediction, postdiction, and perceptual length contraction: a Bayesian low-speed prior captures the cutaneous rabbit and related illusions. Frontiers in Psychology 4:579.
Hassabis D, Kumaran D, Vann SD, Maguire EA (2007) Patients with hippocampal amnesia cannot imagine new experiences. Proc Natl Acad Sci U S A 104:1726-1731.
Ismael J (2007) The Situated Self. Oxford: Oxford University Press.
Kilgard MP, Merzenich MM (1995) Anticipated stimuli across skin. Nature 373:663.
Nunez R, Cooperrider K (2013) The tangle of space and time in human cognition. Trends Cogn Sci 17:220-229.
Penrose R (1989) The emperor's new mind., 1999 Edition. Oxford: Oxford University Press.
Price H (1996) Time's Arrow, Oxford: Oxford University Press.
Race E, Keane MM, Verfaellie M (2011) Medial temporal lobe damage causes deficits in episodic memory and episodic future thinking not attributable to deficits in narrative construction. J Neurosci 31:10262-10269.
Rovelli C (2018) The order of time. New York: Riverhead Books.
Rovelli C (2019) Neither Presentism nor Eternalism. arXiv:191002474
Rovelli C (2020a) Memory and entropy. ArXives : http://arxivorg/abs/200306687.
Rovelli C (2020b) Agency in Physics. ArXives: : http://arxivorg/abs/200705300.
Smolin L (2013) Time reborn: from the crises in physics to the future of the universe. New York: Houghton Mifflin Harcourt.
Suddendorf T, Corballis MC (1997) Mental time travel and the evolution of the human mind. Genet Soc Gen Psychol Monogr 123:133-167.

van Wassenhove V (2009) Minding time in an amodal representational space. Philos Trans R Soc Lond B Biol Sci 364:1815-1830.
Weyl H (2009/1949) Philosophy of mathematics and natural science. Princeton: Princeton University Press.
Zeh, HD (1989) The Physical Basis of the Direction of Time. Berlin: Springer.